\documentclass[%
 reprint,
 amsmath,amssymb,
 aps,
]{revtex4-2}

\usepackage{graphicx}% Include figure files
\usepackage{dcolumn}% Align table columns on decimal point
\usepackage{bm}% bold math
\usepackage{xcolor}

\begin{document}

\title{Measures of Fluctuations for a Liquid Near Critical Drying.}
%\thanks{A footnote to the article title}%

\author{Mary K. Coe}
\author{Robert Evans}
\author{Nigel B. Wilding}
 \affiliation{H.H. Wills Physics Laboratory, University of Bristol, Bristol BS8 1TL, U.K.}%Lines break automatically or can be forced with \\

% \homepage{http://www.Second.institution.edu/~Charlie.Author}

%\date{\today}% It is always \today, today,
             %  but any date may be explicitly specified

\begin{abstract}
We investigate density fluctuations in a liquid close to a solvophobic substrate at which a surface critical drying transition occurs. Using classical density functional theory, we  determine three measures of the spatial extent and strength of the fluctuations, i.e., the local compressiblity $\chi_{\mu}(z)$, the local thermal susceptibility $\chi_T(z)$ and the reduced density $\chi_*(z)$; \textit{z} is the distance from the substrate. Whilst the first measure is frequently used, the second and third were introduced very recently by Eckert et.al., Phys.~Rev.~Lett.~\textbf{125}, 268004 (2020). For state points in the critical drying regime, all three measures, each scaled by its bulk value, exhibit very similar forms and the ratio of $\chi_T(z)$ to $\chi_{\mu}(z)$, for fixed $z$ in the vapour-liquid transition region, is constant. Using a scaling treatment of surface thermodynamics we show that such behaviour is to be expected on general grounds.
\end{abstract}

%\keywords{Suggested keywords}%Use showkeys class option if keyword
                              %display desired
\maketitle

%\tableofcontents

\section{\label{sec:introduction}Introduction}
Recent studies have linked the emergence of vapour-like films near to hydrophobic and solvophobic planar substrates in contact with liquids to an underlying surface phase transition called drying \cite{EvansStewartWilding2016,EvansStewartWilding2017, EvansStewartWilding2019}. Drying is the analogue of the more common wetting where, for sufficiently attractive substrates, a macroscopically thick film of liquid can intrude between the bulk vapour and the substrate. Specifically, the drying transition that occurs when the bulk liquid is at liquid-vapour coexistence and the strength of attractive interactions between the substrate and fluid is lowered, leading to a macroscopically thick intruding vapor film, has been shown to be critical in almost all cases \cite{EvansStewartWilding2019}. In addition, a continuous transition occurs for purely repulsive, or very weakly attractive, substrates as the chemical potential, $\mu$, of an oversaturated liquid is decreased towards the coexistence chemical potential, $\mu_{co}$, e.g. \cite{Dietrich1988, EvansParry1990}. This latter transition is referred to as complete drying.

Continuous transitions are, of course, associated with enhanced density fluctuations. The recent work of Evans et.~al.~\cite{EvansStewartWilding2016,EvansStewartWilding2017} has related the enhanced density fluctuations observed near solvophobic substrates to the critical drying transition. Whilst earlier work on hydrophobic systems also observed an enhancement of density fluctuations \cite{PatelGarde2012,Jamadagni2011,AcharyaGarde2010}, no link was made to critical drying. Many of the earlier studies quantified such fluctuations using the probability distributions of particle occupancy in sub-volumes immediately adjacent to hydrophobic and solvophobic substrates \cite{Jamadagni2011,PatelGarde2012}. However, these measures provide scant information about the spatial extent of the fluctuations, and how their magnitude depends on the strength of substrate-liquid attraction and the deviation from bulk liquid-vapour coexistence, given by $\delta\mu=\mu-\mu_{co}$. Moreover, these earlier studies do not address the important issue of proximity to a critical drying transition. 

Measures which incorporate spatial resolution have typically been termed local compressibilities \cite{AcharyaGarde2010, EvansStewart2015}, and have focused on derivatives of the density profile, $\rho(\mathbf{r})$, of the inhomogeneous fluid. The most natural, and perhaps most widely used, measure is the local compressibility adopted by Stewart and Evans \cite{EvansStewart2015}
\begin{equation}
    \chi_{\mu}(\mathbf{r}) = \left(\frac{\partial \rho(\mathbf{r})}{\partial \mu}\right)_T
    \label{eqn:local_compressibility_definition}
\end{equation}
where the derivative is w.r.t. the chemical potential $\mu$ of the bulk reservoir and the temperature $T$ is fixed. This measure has been utilised in several previous studies of hydrophobic and solvophobic systems, and has been shown to be a powerful indicator of the approach to a critical drying transition \cite{EvansStewartWilding2016,EvansStewartWilding2017}. It provides a much sharper signature of the onset of drying than the density profile itself, i.e. the growth in thickness of a drying film of vapor is much slower than that of the maximum of $\chi_{\mu}(\mathbf{r})$ on approaching the transition. Note that in some earlier papers, for example \cite{EvansParry1990,StewartEvans2012,StewartEvans2014}, $\chi_{\mu}(\mathbf{r})$ was termed the local susceptibility.

Here we enquire whether there are other measures that could provide additional information about the nature of density fluctuations in the inhomogeneous liquid close to critical drying. We were motivated by the recent work of Eckert et. al. \cite{EckertSchmidt2020} who defined two new measures, the local thermal susceptibility $\chi_T(\mathbf{r})$ and the reduced density $\chi_*(\mathbf{r})$, in a similar manner to the local compressibility $\chi_{\mu}(\mathbf{r})$. Specifically, they defined these measures as \cite{EckertSchmidt2020}
\begin{equation}
    \chi_T(\mathbf{r}) = \left(\frac{\partial\rho(\mathbf{r})}{\partial T}\right)_{\mu}
    \label{eqn:local_thermal_susceptibility_definition}
\end{equation}
\begin{equation}
    \chi_*(\mathbf{r}) = \rho(\mathbf{r}) -\mu\chi_{\mu}(\mathbf{r}) -T\chi_T(\mathbf{r})
    \label{eqn:reduced_density_definition}
\end{equation}
Using Grand Canonical Monte Carlo simulations, the authors determined the three quantities, which they term fluctuation profiles, for various model fluids under a variety of confinements, and found that the profiles differed considerably from one other, suggesting they reflect different aspects of the density fluctuations.

At first glance it is not obvious what precisely these quantities measure. Evans et.al. \cite{EvansStewartWilding2017} showed that $\chi_{\mu}(\mathbf{r})$ is the correlator of the total number operator $N$ with the local number density operator at position $\mathbf{r}$. For a planar substrate $\chi_{\mu}(\mathbf{r})$ is an integral of the density-density pair correlation function \cite{EvansStewart2015}. Eckert et.al. \cite{EckertSchmidt2020} identified $\chi_T(\mathbf{r})$ as the correlator (covariance) of the total entropy operator $S$ with the local number density operator and they showed that $\chi_*(\mathbf{r})$ is the difference between the local density and the correlator of the Hamiltonian operator $\mathcal{H}$ with the local number density operator. Near a surface critical transition it is expected that all three quantities should reflect the divergence of the density-density correlation length measured parallel to the substrate. However, the precise nature of the divergence of each quantity is not known and is investigated here.

We use classical DFT to calculate $\chi_{\mu}(z)$ and the new measures $\chi_T(z)$ and $\chi_*(z)$ for a truncated Lennard-Jones (LJ) fluid adsorbed at a planar substrate that exerts an external substrate-fluid potential decaying algebraically with the distance $z$ from the substrate, for state points close to the critical drying transition. For this model system, which is often employed in simulations, it is known that critical drying occurs precisely when the strength of substrate-fluid attraction vanishes, i.e. in the hard-wall limit  \cite{EvansStewartWilding2016,EvansStewartWilding2017}. This allows us to make detailed comparisons between theory and DFT results.
We find that the three measures, each normalized by its value in bulk, exhibit very similar forms in the near critical drying regime. Moreover, at a given $T$, the ratio of $\chi_T(z)$ to $\chi_{\mu}(z)$, measured at fixed $z$ in the edge of the vapour film, is found to be constant in the near critical regime. We explain this observation using i) a scaling treatment of surface thermodynamics and ii) an effective interface potential analysis. Remarkably the result we find from both treatments not only accounts for our DFT results but turns out to be a close analogue of the ratio in bulk where, considering the density of the bulk fluid $\rho_b\equiv\rho_b(\mu,T)$, it follows that
\begin{equation}
    \chi_{T,b} = -\left(\frac{\partial\rho_b}{\partial\mu}\right)_T\left(\frac{\partial\mu}{\partial T}\right)_{\rho_b} \equiv -\chi_{\mu,b}\left(\frac{\partial\mu}{\partial T}\right)_{\rho_b}
    \label{eqn:bulk_relation}
\end{equation}
i.e., the ratio of $\chi_{T,b}$ to $\chi_{\mu,b}$ is the negative of the temperature derivative of $\mu$ at fixed density. Note that $\chi_{\mu,b}=\rho_b^2\kappa_T$, where $\kappa_T$ is the usual (bulk) isothermal compressibility.

Our paper is organized as follows: in Sec. \ref{sec:system_description} we describe the model fluid and substrate, and how these are treated within our microscopic DFT calculations. Sec. \ref{sec:theory} describes the underlying theory, i.e. our treatment of the surface thermodynamics and the pertinent interface potential. We conclude in Sec. \ref{sec:dft_results} with a discussion of our results and their general relevance for fluctuations near interfacial phase transitions.

\section{DFT Calculations for a Model Fluid.}
\label{sec:system_description}
We consider the same model fluid as that treated in \cite{EvansStewartWilding2017}, i.e. a truncated LJ fluid with particles of diameter $\sigma$ and well-depth $\varepsilon$. We use Fundamental Measure Theory \cite{Rosenfeld1989,Roth2010} to treat repulsive interactions between fluid particles, modelling these as hard-spheres and employing the original Rosenfeld hard-sphere functional. Attraction is incorporated using the standard DFT mean field theory treatment \cite{EvansFundInhomFluids}. Specifically, the attractive pair potential used within DFT is given by \cite{EvansStewartWilding2017, EvansFundInhomFluids}
\begin{equation}
    \phi_{att}(r) =
    \begin{cases}
        -\varepsilon & r<r_{min} \\
        4\varepsilon\left[\left(\frac{\sigma}{r}\right)^{12} - \left(\frac{\sigma}{r}\right)^{6}\right] & r_{min}<r<r_{c} \\
        0 & r>r_c
    \end{cases}
    \label{methods:eqn:WCA_LJ_potential}
\end{equation}
where $r$ is the distance between the centres of two fluid particles, $r_{min}=2^{1/6}\sigma$ is the minimum of the pair potential and $r_c=2.5\sigma$ is the cut-off radius of interaction. Within this mean-field approach the bulk critical temperature is $k_{\mathrm{B}}T_c/\varepsilon=1.319442$. This model fluid is in contact with an impenetrable smooth planar substrate composed of particles of diameter $\sigma_s$ of homogeneous density $\rho_s$. The substrate exists in the x-y plane and has its surface located at $z=0$. The external potential exerted by the substrate on a fluid particle takes the form \cite{EvansStewartWilding2017}
\begin{equation}
    V_{ext}(z) = \begin{cases}
        \infty & z<0 \\
        \varepsilon_{sf}\left[\frac{2}{15}\left(\frac{\sigma_s}{z+z_{min}}\right)^9 - \left(\frac{\sigma_s}{z+z_{min}}\right)^3\right] & z>0
    \end{cases}
    \label{methods:eqn:DFT_planar_external_potential}
\end{equation}
where $\varepsilon_{sf}=2\pi\rho_s\varepsilon_s\sigma_s^3/3$ is a measure of the substrate-fluid attraction strength, $\varepsilon_s$ is the well depth of the substrate particle-fluid particle LJ potential, and $z_{min}=(2/5)^{1/6}\sigma_s$ is the location of the minimum of the interaction potential. We set $\sigma_s$ = $\sigma$ and, for numerical reasons, shift the minimum of the potential so that it occurs at the surface of the planar substrate, see \cite{EvansStewartWilding2017}. {The DFT program used within this work can be found at \cite{DFTCode}.

\section{Theory\label{sec:theory}}
\subsection{Surface Thermodynamics and Scaling Argument}\label{sec:theory:thermodynamics}
We consider a system consisting of a smooth planar substrate in contact with a liquid which is at a state point close to its critical drying transition. Due to the symmetry of the substrate, the density profile of such a system varies only along the direction perpendicular to the substrate, $z$, and has the form of a vapour film near to the substrate, with a smooth transition to the bulk liquid near $z= \ell_{eq}$, as sketched in fig. \ref{fig:system_layout}.

\begin{figure}
    \centering
    \includegraphics[clip, trim = 0.0cm 0.5cm 0.5cm 0.7cm, width = 0.45\textwidth]{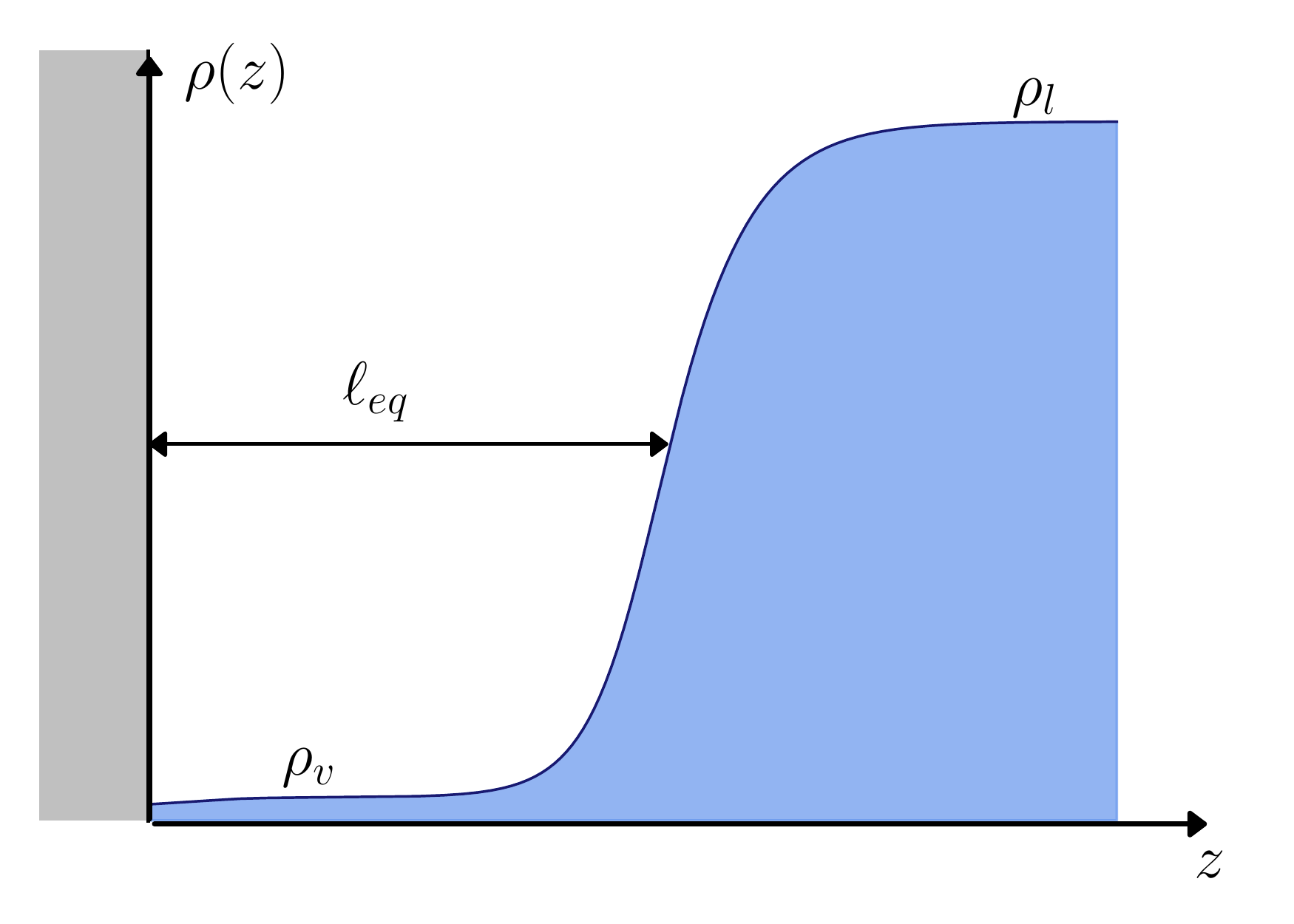}
    \caption{Illustration of the density profile $\rho(z)$ in the near drying situation. A macroscopic planar substrate is in contact with a bulk fluid in its liquid phase. At equilibrium, a film of vapour of thickness $\ell_{eq}$ intrudes between the substrate and liquid. Its extent depends on the  state point, i.e. how close this is to bulk coexistence, measured by $\beta\delta\mu$, and the strength of substrate-fluid attraction, measured by $\varepsilon_s$.}
    \label{fig:system_layout}
\end{figure}

We assume the density profile of such a system can be written as a smooth function, $\rho(z)=S(z-l_{eq})$. Using the definitions given in equations (\ref{eqn:local_compressibility_definition},\ref{eqn:local_thermal_susceptibility_definition}), it follows that
\begin{align}
    \chi_{\mu}(\ell_{eq}) &= -\rho'(\ell_{eq})\left(\frac{\partial \ell_{eq}}{\partial\mu}\right)_T \nonumber\\
    \chi_T(\ell_{eq}) &= -\rho'(\ell_{eq}) \left(\frac{\partial\ell_{eq}}{\partial T}\right)_{\mu}
    \label{eqn:chi_leq_definition}
\end{align}
where $\rho'(\ell_{eq})$ is the spatial derivative of the density profile evaluated at $\ell_{eq}$. Clearly
\begin{equation}
    \frac{\chi_T(\ell_{eq})}{\chi_{\mu}(\ell_{eq})} = \left(\frac{\partial\ell_{eq}}{\partial T}\right)_{\mu}\left(\frac{\partial \ell_{eq}}{\partial\mu}\right)_T^{-1}
    \label{eqn:ratio_leq_version}
\end{equation}
$\ell_{eq}$ can be defined from a thermodynamic quantity, namely the excess Gibbs adsorption, $\Gamma$, as $\ell_{eq}\equiv-\Gamma/A\Delta\rho$ \cite{SullivanGama1986}, where $A$ is the surface area of the substrate-fluid interface and $\Delta\rho=(\rho_l-\rho_v)$ is the difference between the coexisting liquid and vapour densities of the bulk fluid. Eq. (\ref{eqn:ratio_leq_version}) can therefore be rewritten as
\begin{equation}
    \frac{\chi_T(\ell_{eq})}{\chi_{\mu}(\ell_{eq})} = \left(\frac{\partial\Gamma}{\partial T}\right)_{\mu}\left(\frac{\partial \Gamma}{\partial\mu}\right)_T^{-1}
    \label{eqn:rato_adsorption_version}
\end{equation}
Noting that $\Gamma\equiv\Gamma(\delta\mu,T)$, and recalling $\delta\mu=\mu-\mu_{co}$, the first partial derivative on the r.h.s. can be written as 
\begin{equation}
         \left(\frac{\partial\Gamma}{\partial T}\right)_{\mu} = \left(\frac{\partial\Gamma}{\partial T}\right)_{\delta\mu} - \left(\frac{\partial\Gamma}{\partial\delta\mu}\right)_T\left(\frac{\partial\mu_{co}}{\partial T}\right)
         \label{eqn:adsorption_total_derivative}
\end{equation}
It is then important to ascertain which contribution diverges fastest and therefore determines the scaling behaviour close to critical drying. From the standard Gibbs' adsorption equation $\Gamma$ can be written as \cite{Evans1988}
\begin{equation}
    \frac{\Gamma}{A} = -\left(\frac{\partial\gamma_{sl}}{\partial\mu}\right)_T
    \label{eqn:gibbs_adsorption}
\end{equation}
where $\gamma_{sl}$ is the substrate-liquid surface tension, defined as the excess grand potential per unit area. Close to critical drying, the singular part of the tension responsible for the divergence of $\Gamma$, can be written as a scaling function, $\Sigma$, of the variable $\delta\mu/\tilde{t}^{\Delta}$ \cite{Schick1988,Dietrich1988}, where $\tilde{t}=(T-T_D)/T_D$ measures the deviation from $T_D$, the drying temperature for a given choice of fluid-fluid (ff) and substrate-fluid (sf) interaction potentials. $\Delta$ is the (surface) gap exponent, i.e., the analogue of the gap exponent that enters the standard scaling function for the bulk free energy density. The singular contribution to the tension is given by 
\begin{equation}
    \gamma_{sing}  \sim |\tilde{t}|^{2-\alpha_s}\Sigma\left(\frac{\delta\mu}{|\tilde{t}|^{\Delta}}\right)
    \label{eqn:surface_tension_scaling_form}
\end{equation}
where $\alpha_s$ is the surface equivalent of the bulk heat capacity critical exponent. Using eq. (\ref{eqn:gibbs_adsorption}), $\Gamma_{sing}$ can therefore be written as
\begin{equation}
    |\Gamma_{sing}| \sim |\tilde{t}|^{2-\alpha_s-\Delta}\mathcal{L}\left(\frac{\delta\mu}{|\tilde{t}|^{\Delta}}\right)
    \label{eqn:singular_adsorption_scaling_form}
\end{equation}
where $\mathcal{L}$ is the first derivative of $\Sigma$, and is itself a scaling function. The temperature and chemical potential derivatives are then
\begin{align}
    \left(\frac{\partial|\Gamma_{sing}|}{\partial T}\right)_{\delta\mu} &\sim |\tilde{t}|^{1-\alpha_s-\Delta}\mathcal{L}\left(\frac{\delta\mu}{|\tilde{t}|^{\Delta}}\right)  \nonumber \\
    &+ |\tilde{t}|^{1-\alpha_s-\Delta} \left(\frac{\delta\mu}{|\tilde{t}|^{\Delta}}\right)\mathcal{L}'\left(\frac{\delta\mu}{|\tilde{t}|^{\Delta}}\right)  \nonumber \\
    \left(\frac{\partial|\Gamma_{sing}|}{\partial\mu}\right)_T &\sim |\tilde{t}|^{2-\alpha_s-2\Delta}\mathcal{L}'\left(\frac{\delta\mu}{|\tilde{t}|^{\Delta}}\right)
\end{align}
where $\mathcal{L}'$ is the first derivative of $\mathcal{L}$, and the term $(\delta\mu/|\tilde{t}|^{\Delta})\mathcal{L}'(\delta\mu/|\tilde{t}|^{\Delta})$ can be recognised to also be a scaling function. Comparing the exponents, we deduce that, for the second term in eq.~(\ref{eqn:adsorption_total_derivative}) to diverge faster than the first, $\Delta>1$. If this were to be the case, then eq.~(\ref{eqn:rato_adsorption_version}) would become
\begin{equation}
    \frac{\chi_T(\ell_{eq})}{\chi_{\mu}(\ell_{eq})} \sim -\frac{\partial\mu_{co}}{\partial T}
    \label{eqn:ratio_final}
\end{equation}
where it is understood that we are considering the ratio of the fastest diverging contributions. The resulting ratio is simply minus the gradient of the bulk coexistence curve at the given temperature. Note the striking similarity to the corresponding ratio of bulk quantities, eq.~(\ref{eqn:bulk_relation}).

So what is the value of the gap exponent $\Delta$ for a particular system? It is well-known that $\Delta$ satisfies \cite{Schick1988}
\begin{equation}
    \Delta = 2-\alpha_s-\beta_s
    \label{eqn:gap_exponent_equality}
\end{equation}
where $\beta_s$ is the surface critical exponent for the adsorption. $\alpha_s$, $\beta_s$ and therefore $\Delta$ are dependent on the respective ranges of ff and sf interactions \cite{Dietrich1988,Schick1988}. For systems where ff and sf interactions are both long-ranged (LR),  $\alpha_s=-1$ and $\beta_s=-1$ (a linear divergence of the adsorption)\cite{Dietrich1988,Schick1988} which leads to $\Delta=4$. Interactions of this type best describe experimental systems \cite{EvansStewartWilding2019}. Thus, the latter are expected to obey eq.~(\ref{eqn:ratio_final}), implying that all three measures of density fluctuations diverge in the same way on the approach to the critical drying transition.

A more challenging case is that of short-ranged (SR) ff and long-ranged (LR) sf interactions which is pertinent for many simulation studies. {\color{black}In this case, critical drying is not temperature dependent, rather this transition occurs at exactly  $\varepsilon_s=0$ for all temperatures below the bulk critical temperature \cite{EvansStewartWilding2019}. $\varepsilon_s$ therefore acts as the measure of deviation from the drying critical point and takes the role of $\tilde{t}$ in the scaling arguments above. (We return to this important point below.)} For the case of SR ff LR sf interactions, $\alpha_s=1$ and $\beta_s=0$ (a logarithmic divergence of the adsorption) \cite{EvansStewartWilding2017} which leads to a gap exponent  $\Delta=1$. Therefore this is a borderline case for the applicability of eq.~(\ref{eqn:ratio_final}). In the next subsection we argue that such systems do indeed obey eq.~(\ref{eqn:ratio_final}).

\subsection{Interface Potential Analysis}\label{sec:theory:interface_analysis}
In order to ascertain whether the case of SR ff LR sf does obey eq.~(\ref{eqn:ratio_final}) it is necessary to resort to a mesoscopic analysis. Following Evans et. al. \cite{EvansStewartWilding2017}, the excess grand potential per unit surface area of a truncated LJ fluid in contact with a  planar wall can be written as
\begin{equation}
    \omega_{ex}(\ell) = \gamma_{sl} + \gamma_{lv} + \omega_B(\ell) + \delta\mu\Delta\rho\ell\label{eqn:excess_grand_pot}
\end{equation}
where $\Delta\rho$ and $\delta\mu$ are as defined earlier and $\omega_B(\ell)$ is the binding potential, which is the free energy required to bind the liquid-vapour interface to the planar substrate at a distance $\ell$. For a system with SR ff LR sf interactions the binding potential takes the form \cite{EvansStewartWilding2017,EvansStewartWilding2019}
\begin{equation}
    \omega_B(\ell) = a(T)\exp(-\ell/\xi_b) + b(T)\ell^{-2}
    \label{eqn:binding_potential}
\end{equation}
where $\xi_b$ is the correlation length of the bulk vapour (the phase that potentially 'wets' the substrate), $a(T)$ is a positive coefficient, proportional to $\Delta\rho$, with the dimensions of energy per unit area, and $b(T)$ is
\begin{equation}
    b(T) = -b_o\rho_s\varepsilon_s\sigma_s^6
    \label{eqn:b_definition}
\end{equation}
Here, $b_o=\pi\Delta\rho/3$, and $\rho_s$, $\varepsilon_s$ and $\sigma_s$ are as discussed in Sec.~\ref{sec:system_description}. The temperature dependence of both coefficients in eq.~(\ref{eqn:binding_potential}) is determined by that of $\Delta\rho$.

Minimising eq.~(\ref{eqn:excess_grand_pot}) w.r.t. $\ell$ yields the equation determining the equilibrium film thickness:
\begin{equation}
    -\frac{\ell_{eq}}{\xi_b} = \ln\left(\frac{\xi_b}{a(T)}\right) + \ln\left(\delta\mu\Delta\rho - \frac{2b(T)}{\ell_{eq}^3}\right)
    \label{eqn:equilibrium_film_width}
\end{equation}
Clearly critical drying at bulk coexistence, $\delta\mu=0$, occurs at $\varepsilon_{s}= 0$ for all temperatures $T$. This equation therefore predicts $\ell_{eq} \sim -\ln\varepsilon_s + 3\ln\ell_{eq}$, in agreement with Evans et. al. \cite{EvansStewartWilding2017}. {\color{black}Note that the film thickness diverges only in the limit $\varepsilon_s\rightarrow 0$ implying that critical drying occurs for all temperatures below the bulk critical temperature and that $\varepsilon_s$ takes on the role of $\tilde{t}$ within the scaling arguments, as outlined in Sec.\ref{sec:theory:thermodynamics}.} On the approach to complete drying at a hard wall, where $\varepsilon_{s}=0$, we find $\ell_{eq}\sim-\ln\delta\mu$, in agreement with the standard result for complete drying (or wetting) from off-coexistence with SR forces, e.g.~\cite{EvansParry1990} and \cite{Dietrich1988}.

Substituting eq.~(\ref{eqn:equilibrium_film_width}) into eq.~ (\ref{eqn:chi_leq_definition}) yields
\begin{equation}
        \chi_{\mu}(\ell_{eq}) = \xi\Delta\rho\rho'(\ell_{eq})\left(\delta\mu\Delta\rho - \frac{2b}{\ell_{eq}^3}\left(1-\frac{3\xi}{\ell_{eq}}\right)\right)^{-1}
        \label{eqn:chi_mu_leq_SR_ff_LR_sf}
\end{equation}
and
\begin{align}
       \chi_T(\ell_{eq}) &= -\xi\Delta\rho\rho'(\ell_{eq})\left(\delta\mu\Delta\rho - \frac{2b}{\ell_{eq}^3}\left(1-\frac{3\xi}{\ell_{eq}}\right)\right)^{-1} \frac{\partial\mu_{co}}{\partial T} \nonumber \\
       &= -\chi_{\mu}(\ell_{eq})\left(\frac{\partial\mu_{co}}{\partial T}\right)
       \label{eqn:chi_T_leq_SR_ff_LR_sf}
\end{align}
In deriving eq.~(\ref{eqn:chi_T_leq_SR_ff_LR_sf}) we have neglected the temperature dependence of the bulk correlation length $\xi_b$. Including this gives a further, more slowly diverging (logarithmic) contribution. 

Eq.~(\ref{eqn:chi_T_leq_SR_ff_LR_sf}) is in agreement with eq.~(\ref{eqn:ratio_final}); recall we address the ratio of fastest diverging contributions. Hence, we have shown that the borderline case of SR ff LR sf interactions should obey the same relation predicted in the previous subsection that used surface thermodynamics and scaling. We deduce that for SR ff LR sf interactions on the approach to critical drying both $\chi_\mu(\ell_{eq})$ and $\chi_T(\ell_{eq})$ should diverge as $\varepsilon_s^{-1}$. On the approach to complete drying we expect both $\chi_{\mu}(\ell_{eq})$ and $\chi_T(\ell_{eq})$ to diverge as $\sim \delta\mu^{-1}$.

\section{DFT Results \label{sec:dft_results}}
$\chi_{\mu}(z)$ and $\chi_T(z)$ are calculated within DFT by performing numerical derivatives of the density profile with respect to $\mu$ and $T$, respectively. The bulk value $\chi_{\mu,b}$ is obtained by calculating the isothermal compressibility as outlined in Sec. \ref{sec:introduction}, whilst the bulk value $\chi_{T,b}$ is most easily obtained using a numerical derivative. $\chi_*(z)$ and $\chi_{*,b}$ follow directly from eq.~(\ref{eqn:reduced_density_definition}).
\begin{figure}
    \centering
    \includegraphics[width=0.45\textwidth]{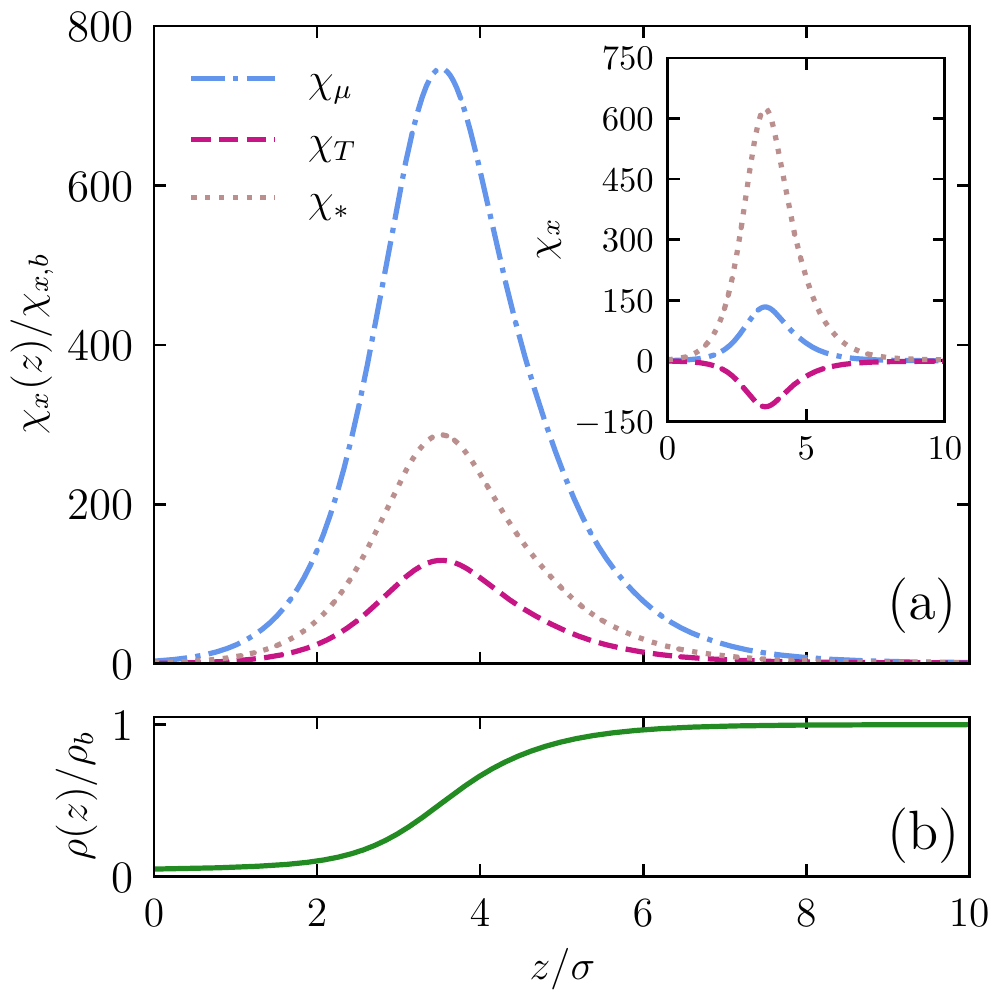}
    \caption{(a) Fluctuation profiles, normalized by their bulk values, for a truncated LJ fluid in contact with a planar hard wall, $\varepsilon_{s}=0.0$. The measure of density fluctuations is denoted as $\chi_{x}$, where $x=\mu,T,*$. Raw fluctuation profiles are given in the inset, and have been made dimensionless by multiplying by $\varepsilon\sigma^{3}$, $\varepsilon\sigma^{3}/k_{\mathrm{B}}$ and $\sigma^{3}$ in the cases of $\chi_{\mu}$, $\chi_T$ and $\chi_*$ respectively. (b) The corresponding density profile, scaled by its bulk value. The state point for the bulk liquid is $T=0.775T_c$ and $\beta\delta\mu=10^{-3}$.}
    \label{fig:profile_comparison}
\end{figure}  

Fig. \ref{fig:profile_comparison}(a) compares fluctuation profiles obtained using DFT for a planar hard wall close to (complete) drying. The chemical potential deviation $\beta\delta\mu=10^{-3}$ is chosen to mimic water at near ambient conditions. Fig. \ref{fig:profile_comparison}(b) shows the density profile of the system for comparison. The fluctuation profiles are scaled by their bulk values in the main plot of fig.~\ref{fig:profile_comparison}(a) and unscaled in the inset. Comparing first the unscaled profiles, it is evident that $\chi_*(z)$ is by far the largest in magnitude , whilst  $\chi_{\mu}(z)$ and $\chi_T(z)$ are similar in magnitude. $\chi_{\mu}(z)$ and $\chi_T(z)$ appear to mirror one another, with the latter taking negative values and the former remaining positive. The differing sign is easily explained by the behaviour of the density profile upon increasing $\mu$ and $T$. When $\mu$ is increased at fixed $T$ the system moves away from bulk coexistence and hence the thickness of the vapour film, measured by $\ell_{eq}$, decreases. The gradient of the density profile in eq.~(\ref{eqn:chi_leq_definition}) is positive and it follows that $\chi_{\mu}(z)$ is also positive. In contrast, increasing $T$ at fixed $\mu$ moves the system closer to liquid-vapour coexistence which increases $\ell_{eq}$ so that $\chi_T(z)$ is negative. (Note that there is no a priori reason for the peak in the $\chi_{\mu}(z)$ and the trough in $\chi_T(z)$ profiles to be perfectly aligned.) Observing the signs of the individual terms in eq.~(\ref{eqn:reduced_density_definition}) it is clear why in the inset of fig. \ref{fig:profile_comparison}(a), $\chi_*(z)$ appears to be the most pronounced measure of density fluctuations near to drying. However, this is somewhat misleading. When the fluctuation profiles are scaled by their bulk values, fig.\ref{fig:profile_comparison}(a) shows that $\chi_{\mu}(z)$ exhibits the most pronounced maximum. All three scaled profiles are positive and their maxima lie very close together, at about 3.5$\sigma$, near where the gradient of the density profile is largest. On decreasing $\beta\delta\mu$ the three (scaled) fluctuation profiles exhibit the same shapes but the position of the maxima increases, consistent with the increase in $\ell_{eq}$, and each peak height increases rapidly, consistent with what we expect from Sec. \ref{sec:theory:interface_analysis} on approaching complete drying. This implies that $\chi_{\mu}(z)$ is the most sensitive measure of density fluctuations in systems near to drying and dictates the behaviour of the other two measures \footnote{In the present analysis we have taken the thermal de Broglie wavelength, $\Lambda$, to be a constant,independent of $T$. This is the usual convention in simulation studies. In \cite{EckertSchmidt2020} the consequences of allowing temperature dependence in $\Lambda$ are addressed. However, when comparing diverging contributions, as we do here, the choice of convention does not matter.}.

We can now examine the reliability of the prediction of eq.~(\ref{eqn:ratio_final}) by using DFT to calculate $\chi_{T}(\ell_{eq})$ and $\chi_{\mu}(\ell_{eq})$ for various $\beta\delta\mu$ and values of $\varepsilon_{sf}$,  proportional to $\varepsilon_s$. $\ell_{eq}$ is calculated from the DFT result for the density profile using $\ell_{eq}\equiv-\Gamma/A\Delta\rho$, i.e. from the measured Gibbs adsorption $\Gamma$. The pertinent ratio is to be compared to $\partial\mu_{co}/\partial T$, calculated from the bulk coexistence curve. The latter is found in the usual way by equating the pressure and chemical potential of the phases. In the case of DFT employing the Rosenfeld functional, the relevant bulk fluid is described by the Percus-Yevick hard sphere model, supplemented by the interparticle attraction of eq.~\ref{methods:eqn:WCA_LJ_potential} treated within mean field  \cite{EvansFundInhomFluids}. For the temperature we consider, $T=0.775T_c$, $(\partial\mu_{co}/\partial T) \approx 0.834$, in reduced units. We choose to define the relative error between the ratio of compressibilities and the gradient of the coexistence curve as
\begin{equation}
    \delta_e=\frac{ \left|\frac{\chi_T(\ell_{eq})}{\chi_{\mu}(\ell_{eq})}+\frac{\partial\mu_{co}}{\partial T}\right|}{\frac{\partial\mu_{co}}{\partial T} }
    \label{eqn:relative_error}
\end{equation}
and plot results for a variety of systems near to critical drying in fig. \ref{fig:error_ratio}. As the limit of critical drying is approached, i.e. $\beta\delta\mu=0,  \varepsilon_{sf}=0$, $\delta_e$ approaches zero, indicating that eq.~(\ref{eqn:ratio_final}) is obeyed. This is displayed very clearly for the two smallest values of $\beta\delta\mu$. The anomalous point, very close to critical drying, with $\beta\delta\mu=10^{-5}$ and  $\varepsilon_{sf}=0$, is related to the {\color{black}numerical} difficulty in evaluating the ratio of local compressibilities in this regime where density fluctuations are especially strong.

\begin{figure}
    \centering
    \includegraphics[width=0.45\textwidth]{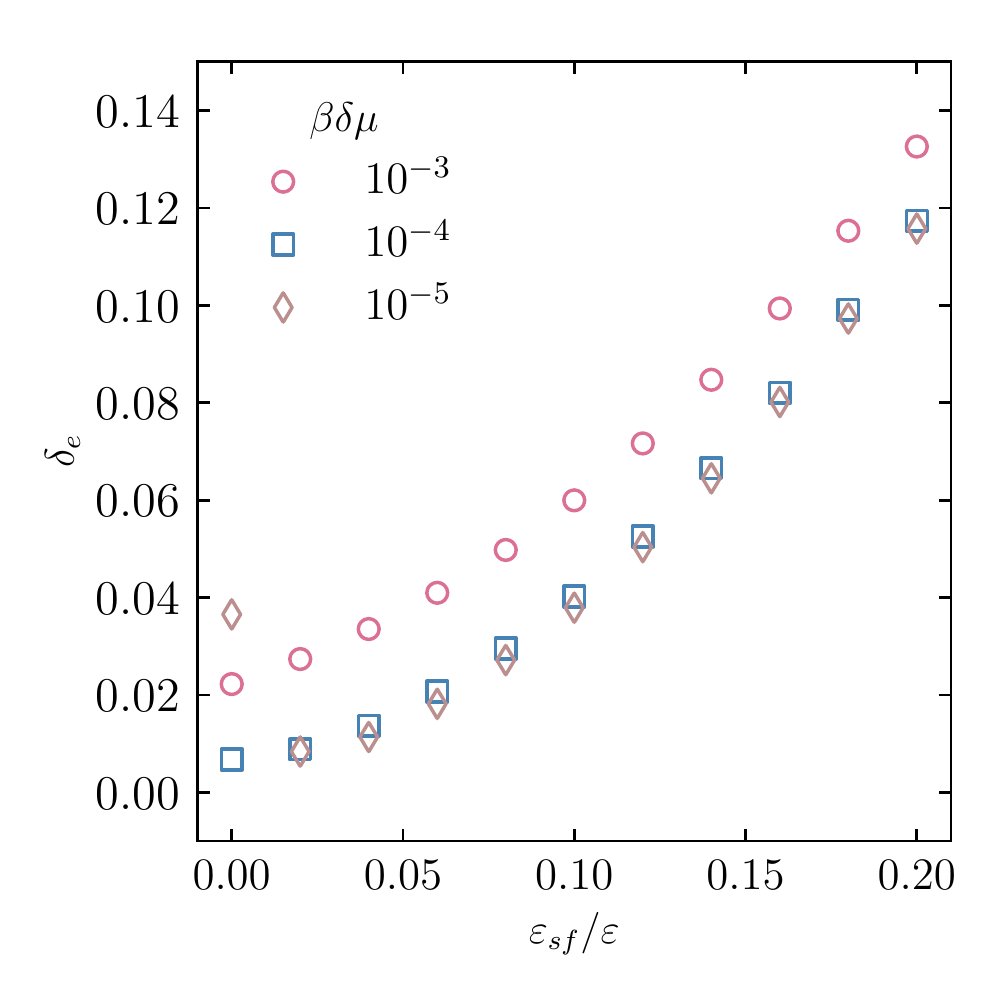}
    \caption{Comparison of the relative error, $\delta_e$, between each side of eq.~(\ref{eqn:ratio_final}) for systems close to critical drying. The dimensionless ratio in the abscissa measures the strength of sf attraction. The temperature was fixed at $T=0.775T_c$.}
    \label{fig:error_ratio}
\end{figure}

\section{Conclusions\label{sec:conclusions}}
We have shown that in the vicinity of a critical drying transition all three measures of density fluctuations introduced by Eckert et.al.~\cite{EckertSchmidt2020} diverge in the same way, i.e. with the same critical exponent. This was demonstrated: i) in Sec. \ref{sec:theory:thermodynamics} by considering the divergence of the local compressibility and thermal susceptibility at $\ell_{eq}$ within a general scaling treatment of surface thermodynamics for a wide class of model systems and ii) in Sec.  \ref{sec:theory:interface_analysis} by considering the form of the same quantities using an interface potential (binding potential) analysis for the borderline case of SR ff and LR sf potentials. DFT results for a system consisting of a (SR) truncated LJ fluid in contact with a planar wall with algebraically decaying (LR) sf attraction were presented in Sec. \ref{sec:dft_results}. These confirm the binding potential predictions and provide explicit results for $\chi_{\mu}(z)$, $\chi_T(z)$ and $\chi_*(z)$, in the vicinity of the critical drying transition. Comparing these three measures of fluctuations, Fig. \ref{fig:profile_comparison} shows that $\chi_{\mu}(z)$ provides the sharpest indicator of the critical drying transition, drives the form of the other measures and therefore can be thought of as the most sensitive measure of density fluctuations in this region. 

Whilst the results presented in Sec. \ref{sec:theory:interface_analysis} and \ref{sec:dft_results} have focused on systems with SR ff LR sf interactions, which are most pertinent to computational studies, we note that the general result of eq.~(\ref{eqn:ratio_final}) is expected to hold for any system with a gap exponent $\Delta>1$. This includes systems with LR ff LR sf interactions, which have $\Delta=4$ \cite{Schick1988,Dietrich1988} and which are pertinent to experimental systems where dispersion forces always prevail at the longest length scales. {\color{black} It also includes systems with SR ff SR sf interactions, where within mean-field, $\alpha_s=0$ and $\beta_s=0$, and hence $\Delta=2$ \cite{Dietrich1988}.} {\color{black} We note that $\chi_{\mu}(z)$ has been calculated exactly for critical wetting in a two dimensional model using transfer matrix methods \cite{Parry1991}. Extending the analysis to $\chi_T(z)$ shows that all measures of density fluctuations exhibit similar forms, however the local compressibility diverges faster than the thermal susceptibility\footnote{{\color{black}We thank a referee for bringing this paper to our attention, and for deriving an expression for $\chi_T(z)$.}}.} Finally, we note that it is possible to derive an expression similar to eq.~(\ref{eqn:ratio_final}) for fluids adsorbed at smooth curved surfaces. In this case, eq.~(\ref{eqn:ratio_final}) acquires an additional curvature dependent term which constrains further the divergence of the local density fluctuations \cite{CoeThesis}.

\begin{acknowledgments}
We thank T. Eckert and M. Schmidt for forwarding  details of their work in advance of publication which motivated our present investigation. R.E. acknowledges Leverhulme Trust Grant EM-2020-029\textbackslash4. This work used the facilities of the Advanced Computing Research Centre, University of Bristol.
\end{acknowledgments}

\appendix

%\nocite{*}

\bibliography{references}% Produces the bibliography via BibTeX.

\end{document}